\documentclass[epj]{svjour}
\usepackage{graphics}
\usepackage{url}
\usepackage{color}
\def\esym{$E_{\rm sym}(\rho)$~}

\def\es0{$E_{\rm sym}(\rho_0)$}

\def\us0{$U_{\rm sym}(\rho_0,k_F)$~}

\def\l0{$L(\rho_0)$~}

\begin{document}
\title{Impact of Symmetry Energy on Sound Speed and Spinodal Decomposition in Dense Neutron-Rich Matter}
\author{Nai-Bo Zhang\inst{1} and Bao-An Li\inst{2}
}                     
\mail{naibozhang@seu.edu.cn; Bao-An.Li@Tamuc.edu}          
\institute{School of Physics, Southeast University, Nanjing 211189, China \and Department of Physics and Astronomy, Texas A$\&$M University-Commerce, Commerce, TX 75429, USA}
\date{Received: date / Revised version: date}
%
\abstract{Using a meta model for nuclear Equation of State (EOS) with its parameters constrained by astrophysical observations and terrestrial nuclear experiments, we examine effects of nuclear EOS especially its symmetry energy \esym term on the speed of sound squared $C^2_s(\rho)$ and the critical density $\rho_t$ where $C^2_s(\rho_t)$ vanishes (indicating the onset of spinodal decomposition) in both dense neutron-rich nucleonic matter relevant for relativistic heavy-ion collisions and the cold $n+p+e+\mu$ matter in neutron stars at $\beta$-equilibrium. Unlike in nucleonic matter with fixed values of the isospin asymmetry $\delta$, in neutron stars with a density dependent isospin profile $\delta(\rho)$ determined consistently by the $\beta$ equilibrium and charge neutrality conditions, the  $C^2_s(\rho)$ almost always show a peak and then vanishes at $\rho_t$. The latter strongly depends on the high-density behavior of \esym if the skewness parameter $J_0$ characterizing the stiffness of high-density symmetric nuclear matter (SNM) EOS is not too far above its currently known most probable value of about $-190$ MeV inferred from recent Bayesian analyses of neutron star observables. Moreover, in the case of having a super-soft \esym that is decreasing with increasing density above about twice the saturation density of nuclear matter, the $\rho_t$ is significantly lower than the density where the \esym vanishes (indicating the onset of isospin-separation instability and pure neutron matter formation) in neutron star cores.
}
\authorrunning{Zhang \& Li}
\titlerunning{Sound Speeds and Spinodal Decompositions in Dense Neutron-Rich Matter}
\maketitle
\section{Introduction}\label{intro}
The energy per nucleon $E(\rho,\delta)$ in cold neutron-rich nucleonic matter of density $\rho$ and isospin asymmetry $\delta=(\rho_n-\rho_p)/\rho$ where $\rho_n$ and $\rho_p$ are the densities of neutrons and protons, respectively, can be written as \cite{Bom91}
\begin{equation}\label{eos0}
E(\rho,\delta)=E_0(\rho)+E_{\rm{sym}}(\rho)\cdot \delta ^{2} +\mathcal{O}(\delta^4)
\end{equation}
where $E_0(\rho)$ is the energy per nucleon in symmetric nuclear matter (SNM) while the symmetry energy $E_{\rm{sym}}(\rho)$ encodes the energy cost to make the matter more neutron-rich \cite{Tesym}. Both the $E_0(\rho)$ and $E_{\rm{sym}}(\rho)$ are important for the stiffness and stability of neutron-rich nuclear matter. For example, within the minimal model of neutron stars consisting of neutrons (n), protons (p), and electrons (e) the incompressibility of n+p+e matter at $\beta$ equilibrium is approximately \cite{Lattimer00,Kubis,JXu2}
$K_{\mu}(\rho)= \rho^2 \frac{d^2 E_0}{d \rho^2} + 2 \rho \frac{dE_0}{d \rho}+\delta^2
[\rho^2 \frac{d^2 E_{\rm sym}}{d \rho^2}
+2 \rho \frac{d E_{\rm sym}}{d \rho} - 2 E^{-1}_{\rm sym}(\rho)
(\rho \frac{d E_{\rm sym}}{d \rho})^2].$
When $K_{\mu}(\rho)$ is negative the corresponding speed of sound $C_s(\rho)$ become imaginary, indicating the onset of the so-called spinodal decomposition. When this happens, the density fluctuations in the dynamically unstable region grow exponentially.

At zero temperature, the $K_{\mu}$ can become negative at both sub-saturation and supra-saturation densities depending on the characteristics of both $E_0(\rho)$ and $E_{\rm sym}(\rho)$.  At low densities, the onset of spinodal decomposition in either the n+p+e matter or purely nucleonic matter is often used to identify the crust-core transition in neutron stars \cite{Lattimer00,Kubis,JXu2} or a liquid-gas phase transition \cite{Bertsch-Siemens,Siemens,Lopez,Goodman,Li-Siemens,Muller} related to the nuclear multifragmentation phenomenon in intermediate energy heavy-ion reactions \cite{Moretto,Bauer,Li-Gross,Xu93,Maria1,Maria2,Pawel20}. In hot dense hadronic matter formed in relativistic heavy-ion reactions, a high-density spinodal decomposition has been associated with a first-order hadron-quark phase transition, see, e.g., Refs. \cite{Risch,Dan98,Dirk95,LiKo,Randrup}. Many potential signatures of such transition have been proposed. For example, the resulting time delay of expansion and/or the rapid growth of fluctuations after the spinodal decomposition may lead to the disappearance of elliptic flow \cite{Dan98} or a minimum in the excitation function of collective flow \cite{Dirk95,LiKo} or an enhanced production of composite particles \cite{Randrup} in relativistic heavy-ion collisions. In cold neutron star matter, when the high-density SNM EOS $E_0(\rho)$ is not very stiff and the symmetry energy $E_{\rm sym}(\rho)$ is very soft, the spinodal decomposition was also found to occur in n+p+e matter at supra-saturation densities in the cores of neutron stars \cite{Kubis,Li20}.  It is thus possible that there are two spinodal instability regions. While the onset of the low-density instability indicates the crust-core transition, the nature of the high-density one is still unclear.
Kubis \cite{Kubis} speculated that the high-density spinodal instability may indicate that the homogeneous phase in the inner core splits into mixed two phases. The new phase may include funny phases or solidification of the central part of stellar core. It was further pointed out that such solid inner core may have important consequences on the rotation of a star and its magnetic properties \cite{Kubis}.

Depending on the EOS of neutron-rich matter \cite{Kubis}, starting from the crust-core transition density the $K_{\mu}$ (and the corresponding $C^2_s$ may first increase and then decrease smoothly to zero at a critical density $\rho_t$ above which the second spinodal instability happens in neutron stars.  In view of the extensive studies in the literature about the speed of sound, here we emphasize two important distinctions of the behavior of $C^2_s$ described above from that studied extensively earlier.
Firstly, often a sudden drop of $C^2_s$ to zero in hadronic matter is taken as an indication of a first-order phase transition. The transition density is model dependent and is often taken as a free parameter \cite{CSS}, see, e.g., Ref. \cite{SHan} for a brief recent review. Thus, a continuously decreasing $C^2_s$ to zero does not signal a first-order phase transition as by definition the latter is characterized by a discontinuous, nonanalytic drop in $C^2_s$. Secondly, there are many interesting studies on the sound speed in different phases of dense matter or neutron stars , see, e.g., Refs. \cite{Ann19,Jerome22,Rho,Malik,Loz,Reddy,Tan,Dris,Agu,Rez,Bra,Kan,Kojo,Mot,Larry,Kumar,Han22,Eck22,Yuki22} and references therein. Model predictions and/or parameterizations of $C^2_s$ \cite{Tews18,Tan2} often purposely
make the $C^2_s$ reach its QCD conformal limit of 1/3 at densities far exceeding that is reachable in neutron stars. As we shall show, depending on the high-density EOS, the possible second spinodal instability in dense neutron-rich matter happens long before reaching the QCD conformal limit. Moreover, some of these studies conjectured that a peak of $C^2_s$ appears at certain density due to some fundamentally new physics happening in the cores of neutron stars, such as the formation of quarkyonic matter \cite{Reddy} and/or derivative contribution from the trace anomaly \cite{Yuki22}.
While we do not question these fundamentally new physics as possible origins of the peak in $C^2_s$, within a traditionally minimal neutron star model, we show that the  $C^2_s(\rho)$ in neutron stars almost always show a peak and then vanishes at certain $\rho_t$ strongly depending on the high-density behavior of \esym if the skewness parameter $J_0$ characterizing the stiffness of high-density SNM EOS is not too far above its currently known most probable value of about $-190$ MeV \cite{Xie-APJ1,Xie-APJ2}.

While effects of nuclear EOS on the crust-core (or liquid-gas) transition density and pressure have been extensively studied, see, e.g., reviews in Ref. \cite{LCK08,Pro14,EPJA-review,Pro21}, little is known about the EOS effects on the onset density of the second spinodal instability in neither heavy-ion reactions nor neutron stars. In this work, we study the $C^2_s(\rho)$ and its possible second vanishing point $\rho_t$ in both dense neutron-rich nucleonic matter relevant for relativistic heavy-ion collisions and the $n+p+e+\mu$ matter in neutron stars at $\beta$-equilibrium and zero temperature. To explore easily the whole EOS parameter space allowed by our current knowledge, we use a meta model for nuclear EOS with its parameters constrained by astrophysical observations and terrestrial nuclear experiments. We focus on effects of nuclear EOS especially its symmetry energy \esym term on the $C^2_s(\rho)$ and $\rho_t$. Because of the long time needed for $\beta$-equilibrium due to weak interactions, it is the partial derivative of pressure $(\partial P(\rho,\delta)/\partial \rho)_\delta$ at a constant $\delta$ determines the $C^2_s$ in relativistic heavy-ion reactions which happen on very short time scales. While in neutron stars at $\beta-$equilibrium, it is the total derivative $dP(\rho,\delta(\rho))/d\rho$ determines the sound speed as the density profile $\delta(\rho)$ of the isospin asymmetry $\delta$ is already determined by the $\beta$-equilibrium and charge neutrality conditions.

We notice that while we can find the second spinodal decomposition onset density $\rho_t$ from the hadronic EOS alone and there are interesting speculations in the literature about what may happen afterwards, in this work we do not study neither the nature of the possible phase transition nor properties of the new phases at higher baryon densities.  Nevertheless, the vanishing point of the speed of sound in dense hadronic matter and its dependence on the hadronic EOS parameters are important information useful for both nuclear physics and astrophysics.

It is also necessary to note that instabilities of nuclear matter with Skyrme forces above saturation density have been discussed previously, see, e.g., Refs. \cite{Kut93,Mar02,Stone03}. Some studies consider such instabilities as being an undesired feature of Skyrme forces in calculating the EOS within the Skyrme-Hartree-Fock approach, see, e.g., Ref. \cite{Mar02,Stone03}, mostly with respect to the desire of supporting neutron stars of masses above about 2M$_{\odot}$ with only the hadronic EOS (plus the EOS of leptons to make the neutron star matter charge neutral and stay at $\beta-$equilibrium) without considering a possible phase transition and including the EOS of the new phase. To our best knowledge, while it may be undesirable in the sense mentioned above, no one has concluded that the hadronic EOS having high-density instabilities is unphysical. In our opinion, the instability just indicates the onset of a new phase requiring additional physics to construct the high-density EOS above the instability onset point. Only with the latter, one can use the astrophysical requirement of supporting neutron stars at least as heavy as the latest most massive neutron star observed. Otherwise, the latter requirement will preclude the EOS in neutron stars from going beyond nucleonic degrees of freedom and the appearance of possible new phases of dense matter. Similarly, it is well known qualitatively that instabilities driven by the onset of Delta-baryons \cite{Lav,Radu}, appearances of hyperons and/or other particles may soften the EOS to levels that can not support 2M$_{\odot}$ neutron stars. Quantitatively, however, the softening of the EOS depends strongly on the poorly known interactions of these new particles with mesons, nucleons and among themselves \cite{Jiang,Cai}. Nevertheless, the inability of supporting 2M$_{\odot}$ neutron stars with the softened EOSs does not necessarily mean the EOS used to find out when/where the instability may occur is wrong, or new phases/particles should not appear. It just calls for new physics to be included once these new phases/particles are produced. 

In the same spirit, in the present study we do not impose the condition that the hadronic EOS itself has to be stiff enough to support 2M$_{\odot}$ neutron stars, although most of the EOS parameters we use are actually within the 68\% confidence boundaries of the posterior probability distribution functions inferred from Bayesian analyses of neutron star observables under the condition that the maximum neutron star mass to be supported by each EOS is a Gaussian function centered at $(2.01\pm 0.04)$ M$_{\odot}$ \cite{Xie-APJ1,Xie-APJ2}. Moreover, we use the minimum model for neutron stars consisting of neutrons, protons, electrons and muons without considering baryon resonances and hyperons. As mentioned earlier, we investigate the behavior of the speed of sound and when/if it reaches zero. Once the spinodal decomposition happens, some new physics has to be incorporated to support 2M$_{\odot}$ neutron stars. For example, consistent with several other studies in the literature, see, e.g., Refs. \cite{Ang,Tang21,Ang1,Jerome-c}, using the same meta model EOS used here it was found that the most probable hadron-quark transition density is $1.6^{+1.2}_{-0.4}\rho_0$ in neutron stars from a Bayesian analysis of available neutron star observables under the condition that the EOS has to be stiff enough to support 2M$_{\odot}$ neutron stars \cite{Xie3}. Such a low hadron-quark transition density seems undesirable as no signature of such low transition density has been seen in heavy-ion collisions, and one may suspect that the low transition density may pose difficulties to support massive neutron stars. But the Bayesian analysis indicates that all one needs is a very stiff EOS of quark matter (namely, high speed of sound in quark matter) and a relatively large skewness parameter $J_0$ of symmetric nuclear matter. Unfortunately, the posterior probability distributions of all nucleonic EOS parameters with or without considering the hadron-quark phase transition in neutron stars are all consistent with our current knowledge from both terrestrial experiments and astrophysical observations, indicating a hadron-quark duality of the EOS/composition in understanding the currently available neutron star observables \cite{Xie3}. In this regard, it is interesting to note that under the assumption that nucleons are the only constituents of neutron star cores, a very recnt Bayesian inference indicates that the current observations are fully compatible with the nucleonic hypothesis \cite{Thi}.
Thus the readers should be advised that all our results and discussions in the following should be understood with the cautions mentioned above. 

The rest of this paper is organized as follows. In the next section, we summarize the major ingredients and available constraints of a meta EOS model for nucleonic matter and neutron star matter. We also recall definitions of a few relevant physical quantities.
We present and discuss our results in section \ref{results}. We then conclude in section \ref{conc}.

\section{Summary of a meta model EOS for neutron-rich matter}
Meta-modeling of nuclear EOS has been widely used in both astrophysics and nuclear physics, see, e.g., Refs. \cite{Margueron:2017eqc,Jerome18,Mondal:2022cva} for detailed discussions on the advantages and disadvantages of such approach.
In our meta model for neutron-rich matter \cite{NBZ18,Zhang:2018bwq}, the SNM EOS $E_0(\rho)$ and symmetry energy $E_{\rm{sym}}(\rho)$ in Eq. (\ref{eos0}) are parameterized respectively according to
\begin{equation}\label{E0-taylor}
  E_{0}(\rho)=E_0(\rho_0)+\frac{K_0}{2}(\frac{\rho-\rho_0}{3\rho_0})^2+\frac{J_0}{6}(\frac{\rho-\rho_0}{3\rho_0})^3,
\end{equation}
where $E_0(\rho_0)$=-15.9 MeV and
\begin{eqnarray}\label{Esym-taylor}
    E_{\rm{sym}}(\rho)&=&E_{\rm{sym}}(\rho_0)+L(\frac{\rho-\rho_0}{3\rho_0})\nonumber\\
    &+&\frac{K_{\rm{sym}}}{2}(\frac{\rho-\rho_0}{3\rho_0})^2
  +\frac{J_{\rm{sym}}}{6}(\frac{\rho-\rho_0}{3\rho_0})^3.
\end{eqnarray}
Such models have been widely used in Bayesian inferences of nuclear EOS from neutron star observables or forward modelings of the structures of neutron stars and nuclei as well as their mergers and collisions.
The parameters involved have the asymptotic boundary conditions that they become Taylor expansion coefficients when the above parameterizations are used near the saturation density $\rho_0$ of SNM and $\delta=0$.
Thus, the fiducial values of the incompressibility $K_0=9\rho_0^2[\partial^2 E_0(\rho)/\partial\rho^2]|_{\rho=\rho_0}$ and skewness $J_0=27\rho_0^3[\partial^3 E_0(\rho)/\partial\rho^3]|_{\rho=\rho_0}$ of SNM EOS $E_0(\rho)$ at $\rho_0$,
the magnitude $E_{\rm{sym}}(\rho_0)$, slope $L=3\rho_0[\partial E_{\rm{sym}}(\rho)/\partial\rho]|_{\rho=\rho_0}$, curvature  $K_{\rm{sym}}=9\rho_0^2[\partial^2 E_{\rm{sym}}(\rho)/\partial\rho^2]|_{\rho=\rho_0}$ and skewness $J_{\rm{sym}}=27\rho_0^3[\partial^3 E_{\rm{sym}}(\rho)/\partial\rho^3]|_{\rho=\rho_0}$ of nuclear symmetry energy $E_{\rm sym}(\rho)$ at $\rho_0$, respectively, from astrophysical observations and terrestrial nuclear experiments provide a guide and some constraints on the parameters of the meta model EOS.

\begin{figure*}[ht]
  \centering
   \resizebox{1.0\textwidth}{!}{
  \includegraphics{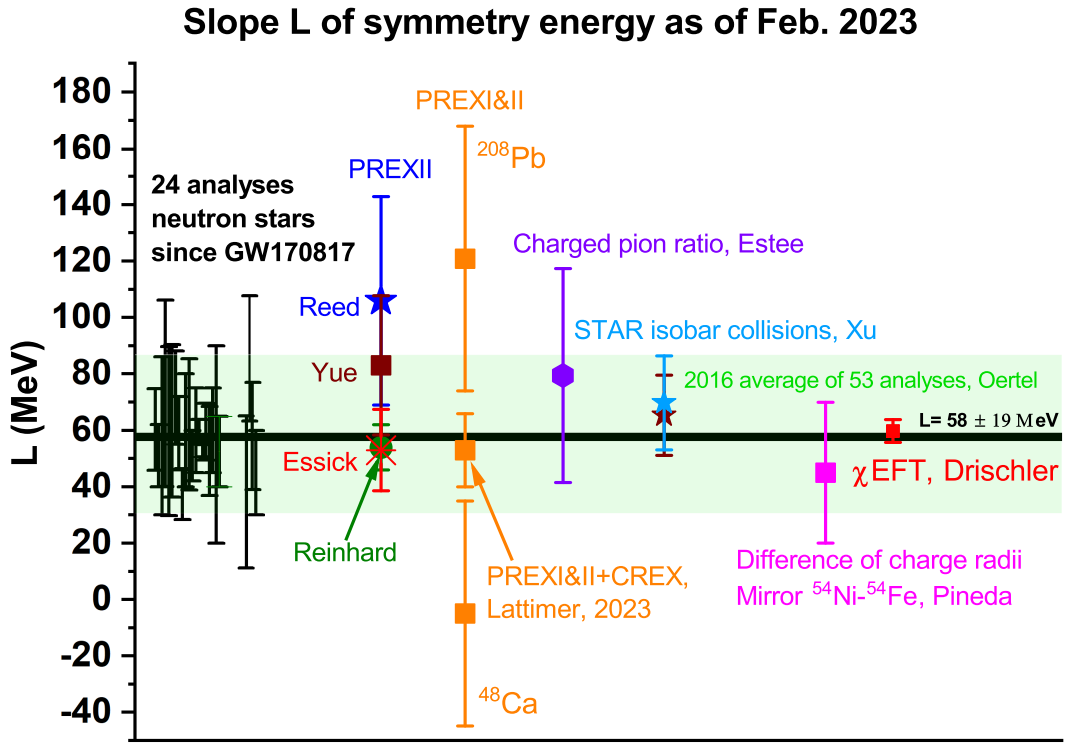}
  }
\vspace{-2cm}
  \caption{An updated compilation of the slope paramter L of nuclear symmetry energy extracted from analyses of several recent terrestrial experimenst in comparison with the 2016 fiducial value of $L=58.7 \pm 28$ MeV (light mint band) \cite{as5} and the average value of $L=58\pm 19$ MeV (black horizonal line) from 24 analyses (listed in Ref. \cite{LiBA21}) of neutron star observables after GW170817. See the text for more details.}\label{Lsym}
\end{figure*}
For studying the high-density spinodal decomposition in this work, we fix the low-density EOS parameters at their currently known most probably values, e.g., $K_0=240$ MeV \cite{Garg18} and $E_{\rm{sym}}(\rho_0)=31.7$ MeV \cite{Li2013} and $L=60\pm 20$ MeV, but vary the high-density parameters within their current uncertain ranges if known.
To justify the $L$ value we use, in Fig. \ref{Lsym} we provide an updated compilation of $L$ with respect to its 2016 fiducial value of $L=58.7 \pm 28$ MeV (light mint band) from averaging results of 53 independent analyses of various terrestrial and astrophysical data available \cite{as5}. Starting from the left are the $L$ values from (1) 24 independent
analyses of neutron star observables carried out by various groups between 2017 and 2021, they give an average $L=58\pm 19$ MeV (horizontal line) \cite{LiBA21}; (2) the original analysis of the PREX-II data \cite{prex2} by Reed et al. \cite{Reed21} and 3 independent analyses of PREX-II data together with different combinations of terrestrial and/or astrophysical data by Reinhard et al. \cite{Reinhard1}, Essick et al. \cite{Essick21} and Yue et al. \cite{Yue22}, respectively; (3) liquid drop model analyses using separately the PREX-I \& II data together, CREX data only, and the combination of all data assuming they are equally reliable by Lattimer \cite{Lattimer23}; (4) charged pion ration in heavy-ion reactions at RIKEN by Estee et al. \cite{Estee21}; (5) the ratio of average transverse momentum (sky blue  star) and the ratio of charged particle multiplicites (black star) in isobar collisions ($^{96}$Zr+$^{96}$Zr and $^{96}$Ru+$^{96}$Ru) from STAR/RHIC experiments analysed by Xu et al. \cite{star}; (6) the difference of charge radii of the mirror pair $^{54}$Ni-$^{54}$Fe by Pineda et al. \cite{mirror}; the chiral EFT prediction by Dirschler et al. \cite{ceft}. It is beyond our ability to comment on the underlying physics differences of the different analyses listed above and the compilation is certainly incomplete as we only used selected results from analyzing some new terrestrial experiments over the last 2 years. However, we notice that the great interest stimulated by the PREX-II result has now been increased even further by the recent publication of the CREX result \cite{crex}. Of course, there are diverse opinions in understanding the PREX and CREX results in the community. For example, on one hand, Ref. \cite{Reinhard2} cautioned that ``Until the tension between theory and experiment, or between the two measurements, is resolved, one should exercise extreme caution when interpreting the new A$_{\rm PV}$ measurements in the context of neutron skins or nuclear symmetry energy". On the other hand, it was found that ``The two experiments separately predict incompatible ranges of $L$ ($L=-5\pm 40$ MeV from CREX and $L = 121\pm 47$ MeV from combined PREXI \& II data, respectively), but accepting both measurements to be equally valid suggests $L = 50\pm 12$ MeV, nearly the range suggested by either nuclear mass measurements or neutron matter theory, and is also consistent
with nuclear dipole polarizability measurements" \cite{Lattimer23}. There are also other efforts to simultaneously explain both PREX-II and CREX data within the same theory/model with very limited success, see, e.g., Refs. \cite{Mondal:2022cva,Zhang22,Tag22,Yuk22,Bis21,Cor22}.
Given the results described above and shown in Fig. \ref{Lsym}, it is seen that the 2016 fiducial value, the average L from the 24 analyses of neutron stars after GW179817 as well as several new results from the last 2 years are all in perfect agreement with the chiral EFT prediction of $L =59.8 \pm 4.1$ MeV \cite{ceft}, albeit with different error bars. Adopting the philosophy that all published experimental data are eqully reliable within their reported confidence boundaries, the value of $L=60\pm 20$ MeV we use in this work is completely consistent with our current best knowledge. Similarly, we notice that the magnitude of symmetry energy $E_{\rm sym}(\rho_0)=31.6\pm 2.7$ MeV \cite{Li2013} or $E_{\rm sym}(\rho_0)=31.7\pm 3.2 $ MeV \cite{as5} based on earlier analyses of mostly terrestrial experiments also agrees very well with the chiral EFT result ($E_{\rm sym}(\rho_0)= 31.7 \pm 1.1$ MeV) \cite{ceft}. These fiducial values have been used already in many of the analyses in extracting the $L$ values.

While several analyses have found strong supports for $K_{\mathrm{sym}}$ to be around $-100 \pm 100$ MeV \cite{Jerome22,Jerome18,LiBA21,Mondal}, a larger range is still often used in the literature. As to the $J_{\rm{sym}}$ parameter, only some theoretical predictions put it in the range between about -200 to +800 MeV \cite{Dut12,Dutra2014,Tews17,Zhang17}. To our best knowledge, none of the analyses of neither astrophysical nor nuclear physics data has provided a reliable constraint on $J_{\rm{sym}}$. Consequently, the high-density behavior of \esym above about $2\rho_0$ is largely open \cite{LiBA21,Zhang21}.

\begin{figure*}[ht]
  \centering
   \resizebox{0.85\textwidth}{!}{
  \includegraphics{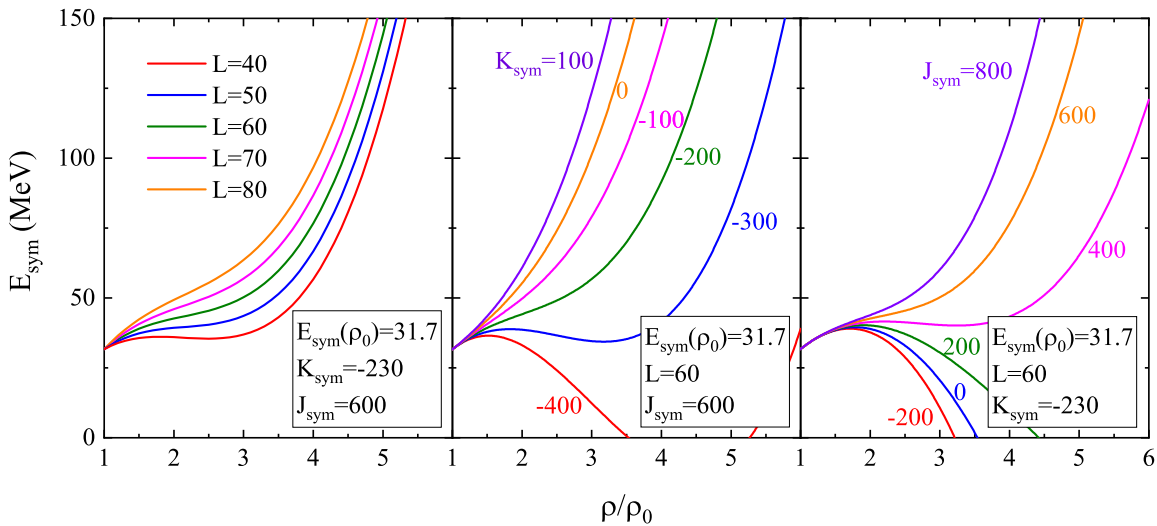}
  }
  \caption{The density dependences of nuclear symmetry energy with different \textcolor[rgb]{0.00,0.07,1.00}{$L$,} $K_{\rm sym}$, and $J_{\rm sym}$ values as indicated.}\label{Esym}
\end{figure*}

To see how diverse the metal model $E_{\rm{sym}}(\rho)$ can be, a few examples are shown in Fig. \ref{Esym} by varying the $L$, $K_{\mathrm{sym}}$, and $J_{\mathrm{sym}}$ parameters in their uncertainty ranges discussed above.
It is seen that the generated $E_{\rm{sym}}(\rho)$ ranges from being super-soft (first increases up to about $2\rho_0$ then start decreasing with increasing density and can even become negative at high densities) to super-stiff (increases with increasing density much fast than linear). The spread of nuclear symmetry energy at suprasaturation densities is consistent with the present constraints \cite{Zhang:2018bwq} and predictions from different models \cite{EPJA-review,LiBA21,Chen18}. Interestingly, one can even simulate a ``cusp" in \esym due to a topology change as predicted by a pseudo-conformal symmetry theory \cite{Rho}.
To our best knowledge, currently neither experimental findings nor fundamental physics principles can rule out any of the high-density behaviors of \esym illustrated in Fig. \ref{Esym}.
It has been recognized that the poorly known isospin-dependent tensor force and the corresponding short-range nucleon-nucleon correlation as well as the spin-isospin dependence of three-body force in neutron-rich matter are among the most important origins of the wide open behavior of high-density symmetry energy \cite{EPJA-review,Cai22}. Impacts of the diverse high-density behaviors of \esym on properties of neutron stars and heavy-ion collisions have also been studied in several works, see, e.g., Refs. \cite{Rho,Kut93,LiBA21,Cai22,Kut94,LiBA02,Kut93a,Szm06,wen,Kho96,Bas07,Ban00,Kubis1}. Is there any physical meaning for the spread in the density dependence of symmetry energy beyond about twice the nuclear saturation density? Yes, albeit there is currently no consensus and model-independent answer, see, e.g., Ref. \cite{EPJA-review} for a recent review. In fact, to pin down the physics origin of the spreading of symmetry energy at suprasaturation densities and find sensitive observables to probe it has been a longstanding goal of many theoretical and experimental efforts. The latest and most detailed discussions on this question can be found in subsection VI-D of a very recent white paper on ``Dense Nuclear Matter Equation of State from Heavy-Ion Collisions" \cite{Sorensen}.

On the other hand, there are several available constraints on the $J_0$ parameter which characterizes the stiffness of SNM EOS at densities above about $(2-3)\rho_0$. For example, recent Bayesian analyses of neutron star radii from LIGO and NICER observations have inferred the most probable value of $J_0$ to be about $J_0=-190_{-40}^{+40}$ at 68\% confidence level \cite{Xie-APJ1,Xie-APJ2}. While its most probable value extracted from a Bayesian analysis of nuclear collective flow in relativistic heavy-ion collisions is $J_0=-180^{+100}_{-110}$ MeV at 68\% confidence level \cite{XieLi-JPG}. Moreover, it was found that to support an earlier reported neutron star of mass 2.17 M$_{\odot}$ a minimum value of about $J_0=-180$ MeV is needed \cite{NBZ217}. While the causality requirement can provide an upper limit on $J_0$,  the range of about $-90 <J_0 < 200$ MeV obtained in this way depends strongly on the correlation between the high-density SNM EOS and symmetry energy \cite{NBZ217}. We will thus explore effects of $J_0$ in a large range around its most probable value of about -190 MeV.

It is worth noting here that most of the microscopic models/theories have a more complex behavior with density compared to the meta model EOS used here. In particular, higher orders in powers of the density and, therefore, the same coefficients may have different meanings in different models/theories. For example, most Skyrme models predict $J_0<-300$ MeV, some RMF models may predict a very negative value as the Skryme forces, or large positive values, and {\it ab-initio} theories predict various negative values. Concerning the $J_{\rm sym}$ all models predict quite large values except for chiral EFT, see, e.g., Refs. \cite{Margueron:2017eqc,Duc} for more detailed discussions. We are not knowledgeable enough to explain why the $J_0$ and $J_{\rm sym}$ from these different models/theories may be very different. In fact, to understand and constrain their predictions in a common framework has been a main science driver in the field. We notice that (1) each one of these models/theories has multiple parameters and they are often strongly correlated, (2) often the structures/formalisms of these models/theories are not flexible enough to be used directly in Bayesian-kind of statistical inferences, (3) one of the main advantages of meta model EOSs is to fully explore the whole EOS parameter space freely without the structural limitations that most microscopic many-body models/theories have, (4) the most probable values and uncertainty ranges of $J_0$ and $J_{\rm sym}$ discussed above are from both Bayesian inferences and bruteforce inversions of neutron star observables using the same meta EOS model used here. Thus, we have at least self-consistency although we can not explain why some other EOS models/theories may give different values for $J_0$ and $J_{\rm sym}$. Considering this situation, we also show as demonstrations in several cases results from using $J_0$ and $J_{\rm sym}$ values significantly different from their most probable values given above.

The energy density in neutron stars consisting of neutrons, protons, electrons, and muons at $\beta$-equilibrium is given by
\begin{equation}\label{lepton-density}
  \epsilon(\rho, \delta)=\rho [E(\rho,\delta)+M_N]+\epsilon_l(\rho, \delta),
\end{equation}
where $M_N$ represents the average nucleon mass and $\epsilon_l(\rho, \delta)$ denotes the lepton energy density.  The pressure is then
\begin{equation}\label{pressure}
  P(\rho, \delta)=\rho^2\frac{d\epsilon(\rho,\delta)/\rho}{d\rho}.
\end{equation}
The pressure in neutron stars at $\beta$-equilibrium becomes barotropic once the density-profile of the isospin asymmetry $\delta(\rho)$ is obtained self-consistently from the $\beta$-equilibrium condition
\begin{equation}\label{equi}
  \mu_n-\mu_p=\mu_e=\mu_\mu\approx4\delta E_{\rm{sym}}(\rho)
\end{equation}
and the charge neutrality condition
$  \rho_p=\rho_e+\rho_\mu.
$
The chemical potential $\mu_i$ for a particle $i$ is obtained from $\mu_i=\partial\epsilon(\rho,\delta)/\partial\rho_i$ \cite{Wiringa88,Lattimer91}. Note that the last term in Eq.~(\ref{equi}) is obtained assuming the mass of electrons can be neglected and the fraction of muons is smaller than that of electrons.
We notice that in nucleonic matter at fixed isospin asymmetries relevant for heavy-ion reactions, the nucleonic part of the above equations gives the corresponding pressure $P(\rho,\delta)$.

For easy of our following discussions, it is useful to recall that the pressure in the n+p+e model of neutron stars is given by \cite{Lat01}
\begin{equation}\label{pre}
P(\rho,\delta)=\rho^2[\frac{dE_{\rm{SNM}}(\rho)}{d\rho}+\frac{dE_{\rm{sym}}(\rho)}{d\rho}\delta^2]
+\frac{1}{2}\delta(1-\delta)\rho E_{\rm sym}(\rho)
\end{equation}
and the isospin profile $\delta(\rho)$ (or the corresponding proton fraction $x_p(\rho)=(1-\delta)/2$)
at $\beta$-equilibrium is determined completely by the \esym via \cite{Lat01}
\begin{eqnarray}\label{xp}
x_p(\rho)= 0.048 \left[E_{\rm sym}(\rho)/E_{\rm sym}(\rho_0)\right]^3
(\rho/\rho_0)(1-2x_p(\rho))^3.
\end{eqnarray}
Based on this equation and as we shall illustrate numerically, as the \esym goes to zero at high densities in the case of a super-soft symmetry energy, the $\delta$ approaches 1 and if the \esym is super-stiff the resulting $\delta$ will vanish at very high densities.

\begin{figure*}[ht]
  \centering
   \resizebox{0.9\textwidth}{!}{
  \includegraphics{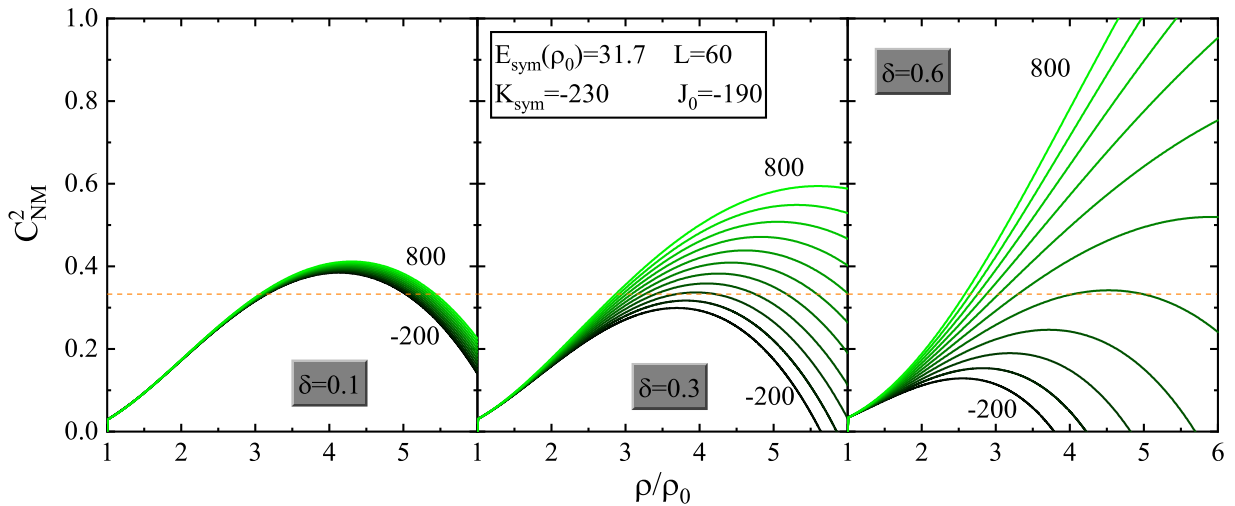}
  }
  \caption{The speed of sound $C^2_{NM}$ in unit of $c^2$ in nucleonic matter with fixed isospin asymmetries as a function of density using the symmetry energy \esym functions with $J_{\rm{sym}}$ varying between -200 and 800 MeV as shown in the right panel of Fig. \ref{Esym}. The orange dashed line corresponds to the conformal limit $C^2<1/3$.} \label{cnm1}
\end{figure*}

As the speed of sound $C_s(\rho)$ is a measure of the stiffness of nuclear EOS, its high-density behavior is largely controlled by the skewness $J_0$ of SNM EOS as well as the curvature $K_{\mathrm{sym}}$ and skewness $J_{\rm{sym}}$ of nuclear symmetry energy. Through the charge neutrality and $\beta$-equilibrium conditions, the $K_{\mathrm{sym}}$ and $J_{\rm{sym}}$ also determine the high-density behavior of the isospin profile $\delta(\rho)$ of neutron stars. As we shall discuss, the $\delta(\rho)$ plays a significant role in determining the density dependence of the sound speed $C_s(\rho)$ and its possible vanishing point $\rho_t$. In the following, we denote the speed of sound squared in nucleonic matter of fixed isospin asymmetry $\delta$ with $C^2_{NM}$ (the so-called adiabatic speed), and that in neutron stars at $\beta$-equilibrium with $C^2_s$ (the so-called equilibrium sound speed). More specifically, they are respectively defined by
\begin{equation}
C^2_{NM}\equiv\left({\partial P\over\partial\epsilon}\right)_{\delta}~{\rm and}~~C^2_s\equiv{dP\over d\epsilon}.
\end{equation}
It has been shown earlier that in terms of the pressure $P$ and incompressibility $K\equiv9\left({\partial P\over\partial\rho}\right)_{\delta}$, the adiabatic sound speed in unit of $c^2$ in nuclear matter can be rewritten as \cite{Jerome22,Blaizot,DongLai}
\begin{equation}\label{cnm}
C^2_{NM}=\frac{K}{9(M_N+E(\rho,\delta)+P/\rho)}.
\end{equation}
Since the density profile $\delta(\rho)$ has to be calculated numerically above the $\mu$ production threshold, no analytical expression for the $C^2_s$ is available.

We notice that besides the equilibrium sound speed one can also calculate the adiabatic sound speed in studying the core g-mode of neutron star oscillations \cite{DongLai,RG1} and its dependence on the \esym \cite{Malik,Loz,EPJA-review,DongLai}. While a fluid element in a neutron star in a core g-mode oscillation is always in mechanical equilibrium with the surroundings, its composition stays a constant during the adiabatic oscillation process because the weak interaction timescale leading to $\beta$-equilibrium is much longer than the dynamical timescale of the oscillation. Namely, the established density profile $\delta(\rho)$ of neutron stars at $\beta$-equilibrium will not change with time during the g-mode oscillation.
As discussed in detail in Ref. \cite{DongLai}, the adiabatic speed of sound is then determined by the partial derivative of pressure with respect to density at a constant proton/electron fraction or isospin asymmetry $\delta$ (the same expression as for $C^2_{NM}$ but with the pressure $P$ for neutron stars). The difference between the adiabatic and the $\beta$-equilibrium sound speeds determines the g-mode frequency. It also determines the so-called Brunt-V\"ais\"al\"a frequency quantifying the stability condition of the core g-mode against convective instability \cite{DongLai,RG1}. Interestingly, the high-density behavior of \esym was found to have a significant effect on the core g-mode frequency, see, e.g., Ref. \cite{DongLai} and the review in Section 7.1 of Ref. \cite{EPJA-review}.

As noticed earlier, there are many interesting studies on the sound speed in different phases of dense matter or neutron stars under various conditions, see, e.g., Refs. \cite{Ann19,Jerome22,Rho,Malik,Loz,Reddy,Tan,Dris,Agu,Rez,Bra,Kan,Kojo,Mot,Larry,Kumar,Han22,Eck22,Yuki22}. As outlined above,  we focus on studying the density dependence of the sound speed in purely nucleonic matter or the $n+p+e+\mu$ matter and its possible vanishing point $\rho_t$. The latter may be used as an input in constructing more complicated EOSs to study some of the interesting and unresolved issues about the phase diagram of dense neutron-rich matter.

\section{Results and Discussions}\label{results}
In the following, we present our results on the speeds of sound in nucleonic matter and neutron star matter separately. We examine effects of the EOS parameters on the density dependence of the speeds of sound and their possible vanishing points.

\subsection{Speed of sound in neutron-rich nucleonic matter with fixed isospin asymmetries}
Shown in Fig. \ref{cnm1} is the speed of sound $C^2_{NM}$ in nucleonic matter with fixed isospin asymmetries of $\delta=0.1, 0.3$ and 0.6, respectively, as a function of density using the symmetry energy \esym functions with $J_{\rm{sym}}$ varying between -200 and 800 MeV as shown in the right panel of Fig. \ref{Esym}. All other EOS parameters are fixed at their currently known most probable values as we discussed earlier. As one expects, as the matter becomes more neutron-rich, effects of $J_{\rm{sym}}$ become larger.
Interestingly, with relatively small $\delta$ and/or $J_{\rm{sym}}$ values the $C^2_{NM}$ shows a peak. From inspecting the expression of Eq. (\ref{cnm}) for  $C^2_{NM}$ in the limiting case of SNM, one can see that while the denominator keeps increasing with density, the incompressibility $K$ in the numerator peaks at certain density since the $K_0$ is positive but $J_0$ is negative. Thus, the $C^2_{NM}$ in SNM normally has a peak. With the parameters $L$ and $K_{\rm{sym}}$ fixed, the isospin asymmetric energy, pressure, and its derivative all increase with positive and increasing $J_{\rm{sym}}$. Their net effect is adding a positive and continuously increasing contribution to $C^2_{NM}$, making its peak disappears gradually when the \esym becomes very stiff for large $J_{\rm{sym}}$ values. Of course, if one uses negative $J_{\rm{sym}}$ values leading to super-soft \esym functions, the $C^2_{NM}$ not only has a peak but also vanishes at a lower density $\rho_t$ as we shall discuss next.

\begin{figure}[ht]
  \centering
   \resizebox{0.48\textwidth}{!}{
  \includegraphics{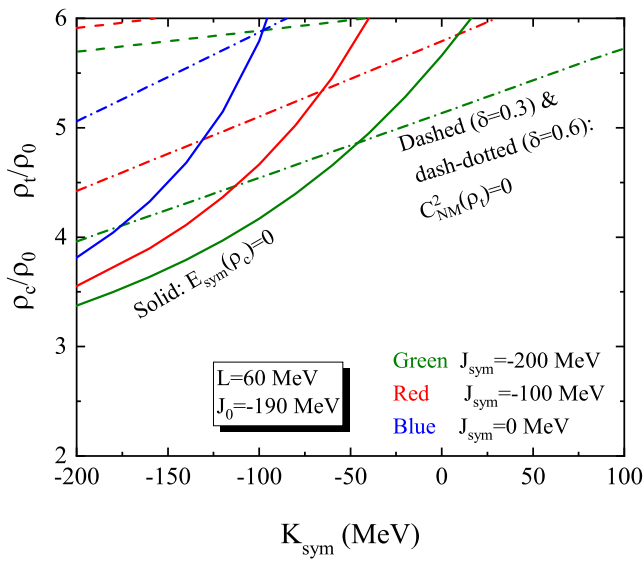}
  }
  \caption{A comparison of the transition density $\rho_t$ where $C^2_{NM}(\rho_t)$=0 and the critical density $\rho_c$ where $E_{\rm sym}(\rho_c)=0$ indicating the onset of isospin separation instability with EOS parameters leading to super-soft symmetry energies. }\label{rhotc}
\end{figure}
As shown in Fig. \ref{Esym}, with small or negative $J_{\rm{sym}}$ values, the \esym is super-soft and vanishes at some point or becomes negative at higher densities. It is thus interesting to compare the transition density $\rho_t$ where $C^2_{NM}(\rho_t)$=0 and the critical density $\rho_c$ where $E_{\rm sym}(\rho_c)=0$.  It is known that a vanishing \esym in nucleonic matter indicates the onset of the so-called isospin separation instability, namely it is energetically more favorable to split SNM into pure neutron matter and proton matter when the \esym becomes negative \cite{Kut94,LiBA02}. For neutron stars, two isospin separation mechanisms were considered by Kutschera et al. \cite{Kut94}: A bulk separatmn of protons and neutrons or formation of a neutron bubble around a single localized proton (the so-called proton polaron). The latter was considered to be more likely to occur than the bulk separtion in neutron star matter. Moreover, it was speculated that the proton polarons in a neutron star core could form a lattice due to the Coulomb interaction at low temperature. Properties of such matter has been studied in a series of papers, see, e.g., Refs. \cite{Kut93,Kut93a,Szm06}. Possible ramifications of isospin separation instability in heavy-ion reactions and neutron stars have also been studied in several other works, see., e.g., Refs. \cite{Rho,LiBA21,Cai22,LiBA02,wen,Kho96,Bas07,Ban00,Kubis1}. However, many interesting issues remain to be addressed.

\begin{figure}[ht]
  \centering
   \resizebox{0.48\textwidth}{!}{
  \includegraphics{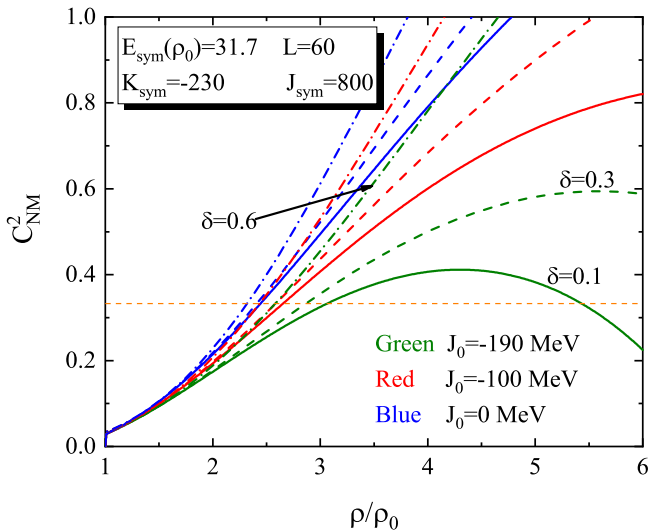}
  }
  \caption{Effects of $J_0$ characterizing the stiffness of SNM EOS on the speed of sound in nucleonic matter with the \esym parameters fixed at the values specified leading to a stiff \esym function. The solid, dashed and dot-dashed lines are for $\delta=0.1, 0.3$ and 0.6, respectively. The orange dashed line corresponds to the conformal limit $C^2<1/3$.} \label{J0effect}
\end{figure}

Shown in Fig. \ref{rhotc} is a comparison of the two transition densities where either $C^2_{NM}(\rho_t)$=0 or $E_{\rm sym}(\rho_c)=0$ as a function of $K_{\rm{sym}}$ with several small $J_{\rm{sym}}$ values all leading to super-soft \esym functions. As illustrated in Fig. \ref{Esym}, it is the combination of $K_{\rm{sym}}$ and $J_{\rm{sym}}$ that is determining the high-density behavior of \esym above about $2\rho_0$. It is seen that when both the $K_{\rm{sym}}$ and $J_{\rm{sym}}$ are small leading to very soft \esym functions that vanish quickly, the critical density for $E_{\rm sym}(\rho_c)=0$ (isospin separation instability) is smaller than that for $C^2_{NM}(\rho_t)$=0 (spinodal decomposition instability). Since most of the existing analyses support $K_{\rm{sym}}\approx -100\pm 100$ MeV at 68\% confidence level and if indeed the $J_{\rm{sym}}$ is small, a super-soft \esym is realized, and then the isospin separation occurs first. However, as we discussed earlier, there is currently no experimental constraint on the $J_{\rm{sym}}$, leaving all possibilities open. Thus, our findings here add more importance for constraining the high-density behavior of nuclear symmetry energy especially its skewness parameter $J_{\rm{sym}}$.

In the results presented above, we have used the currently known most probable value of $-190$ MeV for the skewness parameter $J_0$ of SNM EOS. As we discussed earlier, the $J_0$ still has a large uncertainty range. To reveal its quantitative effects on the speed of sound, we vary $J_0$ between 0 and -190 MeV using a relatively stiff \esym function created with $K_{\rm{sym}}=-230$ MeV and $J_{\rm{sym}}=800$ MeV. The results are shown in Fig. \ref{J0effect}. It is seen that the peak in the speed of sound gradually disappears if the $J_0$ is far from its most probable value of -190 MeV such that the SNM EOS becomes very stiff at high densities. This effect is stronger for more neutron-rich matter as a very large $J_{\rm{sym}}$ of 800 MeV is used. These features are easily understood from the terms involved in the expression for $C^2_{NM}$ in Eq. (\ref{cnm}).

In short, in nucleonic matter with fixed isospin asymmetries, if the $J_0$ is close to its most probable value of -190 MeV and the \esym is not too stiff, the $C^2_{NM}(\rho)$ shows a peak and may become zero at high densities. However, if the \esym is super-stiff especially with high isospin asymmetries the $C^2_{NM}(\rho)$ will keep increasing with density until reaching the causality limit. As we shall illustrate and discuss next, the physical consequence of the last case will be changed completely when the $\beta$-equilibrium condition is introduced in building the EOS for neutron stars.

\subsection{Speed of sound in charge neutral $npe\mu$ matter in neutron stars at $\beta$-equilibrium}
Besides the contributions from leptons, the most important difference between the EOSs of nucleonic matter at fixed $\delta$ and neutron star matter is the consequence of charge neutrality and $\beta$-equilibrium in the latter. It is well known that in neutron stars at $\beta$-equilibrium, the isospin profile $\delta(\rho)$ is uniquely determined by the Eq.(\ref{equi}). As mentioned earlier, the resulting $\delta(\rho)$ approaches one (zero) when the \esym becomes zero (very high or stiff). This result is also easy to understand qualitatively from the energy conservation point of view. As shown in Eq. (\ref{Esym}), the $E(\rho,\delta)\propto E_{\rm{sym}}(\rho )\cdot \delta ^{2}$. To conserve the total energy, if the $E_{\rm{sym}}(\rho)$ is high (low) at certain density $\rho$, the $\delta(\rho)$ there will be low (high).  In both cases, the same isospin asymmetry $\delta$ of the whole system is just being distributed to different density regions according to the well-known isospin fractionation mechanism in asymmetric nuclear matter \cite{Muller,liko,baran,shi,HXu}. Considering two connected regions with local densities $\rho_1$ and  $\rho_2$, respectively, the chemical equilibrium condition requires that \cite{shi} $E_{\rm sym}(\rho_1)\delta(\rho_1)=E_{\rm sym}(\rho_2)\delta(\rho_2)$. Thus, for a given \esym function, the local isospin asymmetries will adjust themselves according to the relative symmetry energy in those regions.

\begin{figure}[ht]
\vspace{-0.5cm}
  \centering
   \resizebox{0.5\textwidth}{!}{
  \includegraphics{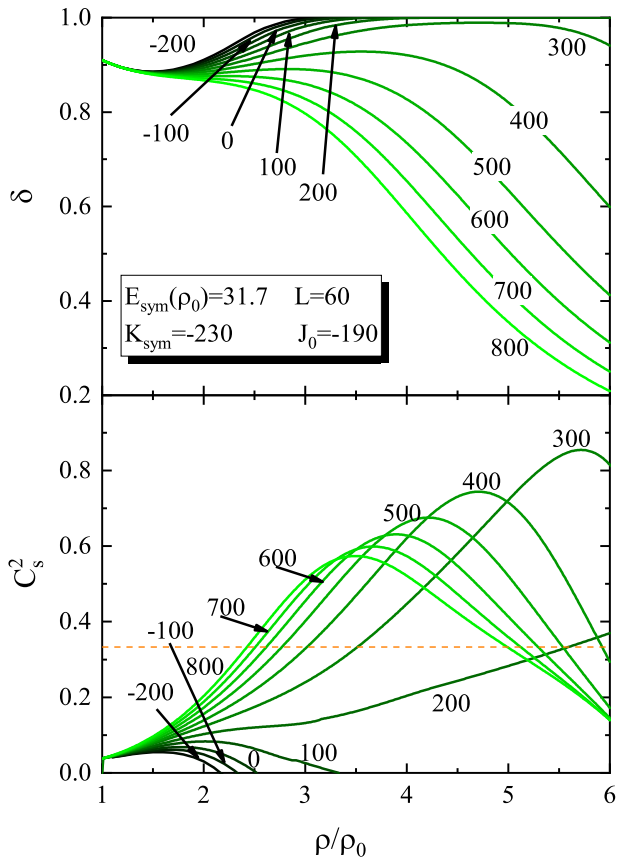}
  }
  \vspace{-1cm}
  \caption{The density profile of isospin asymmetry $\delta(\rho)$ (upper panel) and the corresponding sound speed squared $C^2_s(\rho)$ in unit of $c^2$ in neutron stars at $\beta$-equilibrium with $J_{\rm{sym}}$ varying between -200 and 800 MeV but other parameters fixed at their currently known most probable values indicated. The orange dashed line corresponds to the conformal limit $C^2<1/3$.}\label{cs2-ns}
\end{figure}
Shown in Fig. \ref{cs2-ns} are the density profile of isospin asymmetry $\delta(\rho)$ (upper panel) and the corresponding squared speed of sound in neutron stars at $\beta$-equilibrium with $J_{\rm{sym}}$ varying between -200 and 800 MeV but other parameters fixed at their currently known most probable values. Again, the corresponding \esym is the one shown in the right panel of Fig. \ref{Esym}. Indeed, it is seen that the super-stiff \esym with $J_{\rm{sym}}$=800 MeV leads to a vanishing $\delta$ (SNM) at high densities. While the super-soft \esym with $J_{\rm{sym}}\leq$0 leads quickly to $\delta=1$ (pure neutron matter) at high densities. Namely, in both cases the product of \esym and $\delta$ at high densities reach zero as required by the $\beta$-equilibrium and charge neutrality conditions. Consequently, regardless of the high-density behavior of \esym (super-soft or super-stiff), the equilibrium speed of sound $C^2_s(\rho)$ in neutron stars at $\beta$-equilibrium always show a peak with its position depends on the values of $J_{\rm{sym}}$ and other EOS parameters as we shall discuss in more detail. Therefore, compared to the results shown in the previous subsection, it is clearly seen that the high-density behaviors of sound speeds in neutron stars at $\beta$-equilibrium and nucleonic matter with fixed $\delta$ values are very different.
\begin{figure}[ht]
  \centering
   \resizebox{0.5\textwidth}{!}{
  \includegraphics{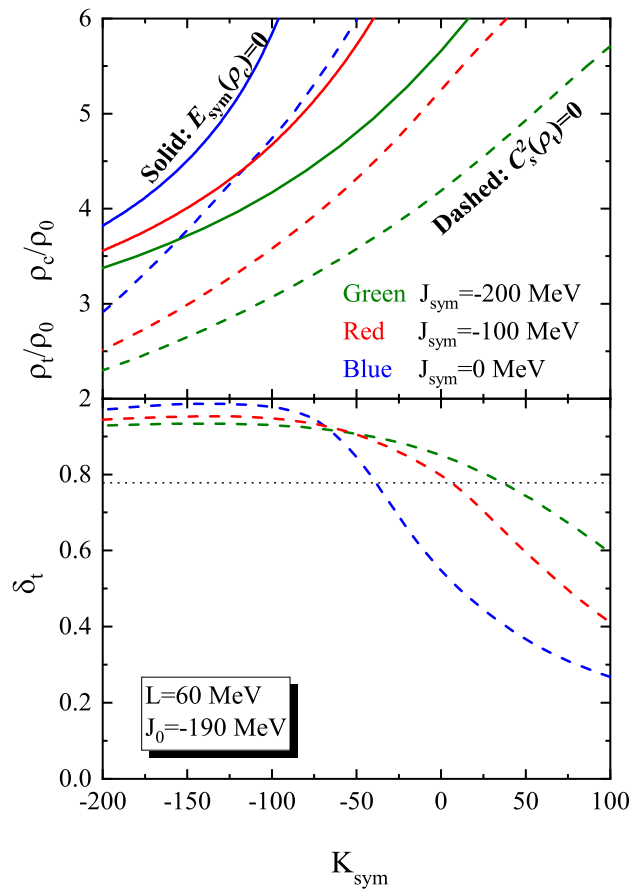}
  }
  \caption{Upper panel: A comparison of the critical ($\rho_c/\rho_0$) and thansition ($\rho_t/\rho_0$) densities where the symmetry energy $E_{\rm sym}(\rho_c)=0$ with that where the sound speed $C^2_{s}(\rho_t)=0$ in neutron stars at $\beta-$equilibrium as functions of $K_{\rm sym}$ with $J_{\rm sym}=0, -100$ and -200 MeV, respectively. Lower panel: The transition asymmetry $\delta_t$ at $\rho_t$ where $C^2_{s}(\rho_t)=0$ as functions of $K_{\rm sym}$ with $J_{\rm sym}=0, -100$ and -200 MeV. The horizontal dotted line corresponds to the critical $\delta$ below which the direct URCA process happens. All other parameters are fixed at their currently known most probable values.}\label{stability}
\end{figure}

Again, it is interesting to compare the critical density $\rho_c/\rho_0$ where the symmetry energy $E_{\rm sym}(\rho_c)=0$ with $\rho_t/\rho_0$ where the sound speed $C^2_{s}(\rho_t)=0$ in the case of having super-soft \esym functions. Shown in the upper panel of Fig. \ref{stability} are the two critical densities as functions of $K_{\rm sym}$ with $J_{\rm sym}=0, -100$ and -200 MeV, respectively, and all other parameters fixed at their currently known most probable values. The horizontal dotted line corresponds to the critical $\delta_t$ below which the direct URCA process happens. Several interesting observations and speculations can be made: (1) since the $\delta_t$ becomes less than the direct URCA limit mostly with large/positive $K_{\rm sym}$ values far above its most probable value, the direct URCA is unlikely to happen before the spinodal decomposition sets in. To our best knowledge, currently there is no strong observational evidence for direct URCA and most of the cooling curves can be well described by the slow modified URCA processes within normally very large error bars. We thus can not say much about the implication of our work here on cooling of neutron stars, (2) in the whole EOS parameter space considered, the spinodal decomposition onset density $\rho_t$ is always lower than the critical density $\rho_c$ where $E_{\rm sym}(\rho_c)=0$ when the isospin separation instability sets in. Around the most probable value of $K_{\rm sym}=-100$ MeV, as seen in the upper panel of Fig. \ref{stability}, the spinodal decomposition density is around the outer core density while the isospin separation density is around the inner core density. Thus, we speculate that these two instabilities may happen in the outer and inner core regions of neutron stars separately.
Consequences and/or observational evidences of this speculation deserves further studies. We emphasize again that the isospin separation instability only happens when the high-density symmetry energy is super-soft. On the other hand, regardless of the high-density behavior (super-soft or super-stiff) of symmetry energy, the speed of sound almost always first goes up, then goes down and finally vanishes at $\rho_t$ in neutron stars.  Of course, as we emphasized earlier, at this time we do not know what new phases may appear and how to handle them above $\rho_t$.

\begin{figure}[ht]
  \centering
   \resizebox{0.55\textwidth}{!}{
  \includegraphics{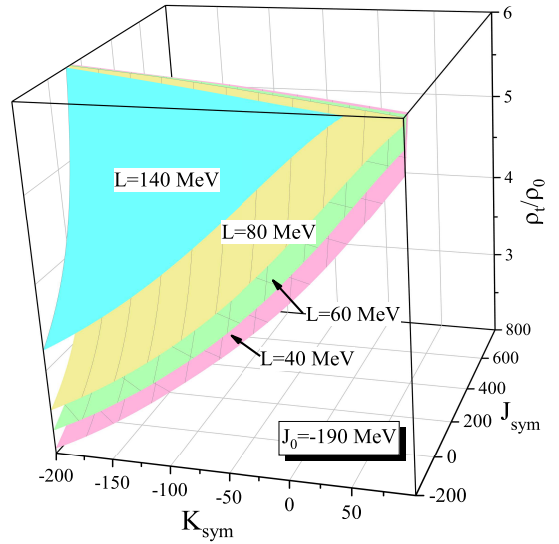}
  }
  \caption{The spinodal decomposition onset density $\rho_t$ in 3D in neutron stars at $\beta-$equilibrium as a function of $J_{\rm sym}$ and $K_{\rm sym}$ with $L=$ 40, 60, 80 and 140 MeV, respectively. All other parameters are fixed at their currently known most probable values.}\label{3D}
\end{figure}
\begin{figure*}[ht]
  \centering
   \resizebox{0.9\textwidth}{!}{
\includegraphics{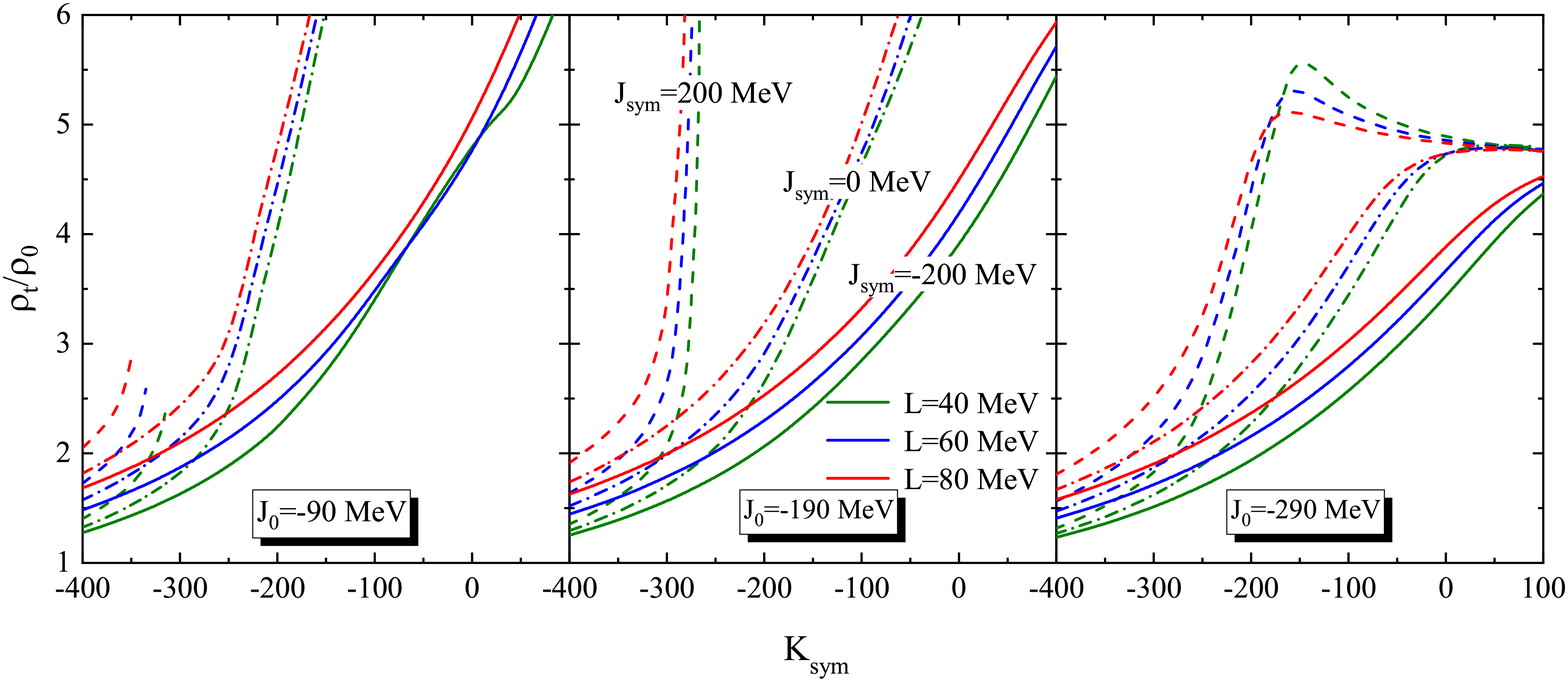}
  }
  \caption{Effects of the skewness $J_0$ of SNM EOS on the spinodal decomposition onset density $\rho_t$ in neutron stars at $\beta$-equilibrium.}\label{J0rt}
\end{figure*}
\subsection{Onset density $\rho_t$ of spinodal decomposition in neutron stars at $\beta$-equilibrium}
We now focus on examining the spinodal decomposition onset density $\rho_t$ where $C^2_{s}(\rho_t)=0$ in neutron stars at $\beta$-equilibrium. First, we fix the $J_0$ at its most probable value of -190 MeV and study the dependence of $\rho_t$ on the slope $L$, curvature $K_{\rm sym}$, and skewness $J_{\rm sym}$ parameters of \esym within their current uncertainty ranges in Fig. \ref{3D}. Overall, a softer \esym leads to a smaller $\rho_t$. Approaching the lower limits of $K_{\rm sym}$ and $J_{\rm sym}$ where the \esym is super-soft, the $\rho_t$ can be as low as about $2\rho_0$ depending on the L value used. Interestingly, such a low onset density of spinodal decomposition is actually consistent with the lower boundary of hadron-quark transition density inferred from several recent Bayesian analyses of neutron star observables \cite{Ang,Tang21,Jerome-c,Xie3,Ferr}. However, at the other limit where both the $K_{\rm sym}$ and $J_{\rm sym}$ take large positive values, the \esym is super-stiff and the resulting $\rho_t$ are larger than $6\rho_0$. It is also seen that varying the $L$ parameter from 40 to 80 MeV causes an approximately 30\% change in $\rho_t$. The PREX-II result indicates $L=106\pm 37$ MeV \cite{Reed21}. While its mean is far above the fiducial value of L as shown in Fig. \ref{Lsym}, its lower boundary is in the 2016 fiducial band of L. To see implications of its upper boundary on the transition density $\rho_t$, results with $L=140$ MeV is also shown in Fig. \ref{3D}. Obviously, it makes the symmetry energy very stiff with the same $K_{\rm sym}$ and $J_{\rm sym}$ parameters compared to the results using lower L values. Consequently, the  $\rho_t$ becomes much higher as one expects. The wide range of $\rho_t$ due to the uncertainty of the \esym especially at high densities signifies again the need to better constrain the density dependence of nuclear symmetry energy.

Next, we examine in Fig. \ref{J0rt} effects of $J_0$ on $\rho_t$ while the \esym parameters are also varied around their most probable values. For exploration purposes only, in one of the calculations we purposely used a very low $J_0$ of -290 MeV leading to an EOS that is not stiff enough to support a neutron star of mass 2.01 M$_{\odot}$. We also extended the lower limit of $K_{\rm sym}$ to -400 MeV that is far below its $1\sigma$ lower boundary of $K_{\rm sym}\approx -100 \pm 100$ MeV from several surveys of its constraints in the literature \cite{LiBA21}. It is seen that even with such a low $K_{\rm sym}$ (which controls the behavior of \esym around $(1-2)\rho_0$) value, the spinodal decomposition may still happen at  densities below about $(1.5-3)\rho_0$. However, as the $K_{\rm sym}$ increases in the case of using a very stiff SNM EOS (e.g., with $J_0=-90$ MeV shown in the left panel far above its most likely value of $-190$ MeV used in the middle panel) and a large $J_{\rm sym}$, it may not occur as indicated by the ending of the $\rho_t$ curves with $J_{\rm sym}$=200 MeV in the left panel.

At the other extreme of using a very soft SNM EOS with $J_0=-290$ MeV shown in the right panel), the $\rho_t$ is again dominated by the SNM EOS and becomes less dependent on the \esym as indicated by the tendency of saturation at the high $K_{\rm sym}$ limit in the right panel in Fig. \ref{J0rt}. These features indicate that there are strong interplays and correlations among the SNM EOS and \esym parameters as one expects.

In short, unlike in nucleonic matter with fixed $\delta$ values, the speed of sound in neutron stars at $\beta$-equilibrium almost always show a peak as long as the $J_0$ characterizing the stiffness of high-density SNM EOS is not too far above its currently known most probable value of about $-190$ MeV. Moreover, the spinodal decomposition onset density $\rho_t$ strongly depends on the high-density behavior of nuclear symmetry energy $E_{\rm sym}(\rho)$.

\section{Summary}\label{conc}
In summary, within a meta EOS model we studied the speed of sound and the possible onset of high-density spinodal decomposition in both nucleonic matter with fixed isospin asymmetries and the $n+p+e+\mu$ matter in neutron stars at $\beta$-equilibrium. While in nucleonic matter with fixed isospin asymmetries the speed of sound may continuously increase with density and does not undergo a spinodal decomposition especially for neutron-rich matter if the \esym is very stiff, the sound speed in neutron star matter at $\beta$-equilibrium almost always show a peak at certain density depending strongly on the high-density behavior of \esym if the skewness $J_0$ of SNM EOS is not too far above its currently known most probable vale of about $-190$ MeV. Moreover, the spinodal decomposition onset density depends sensitively on the high-density behavior of nuclear symmetry energy $E_{\rm sym}(\rho)$. Furthermore, in the case of using a super-soft $E_{\rm sym}(\rho)$, the spinodal decomposition occurs at a density lower than the critical density above which isospin separation instability occurs.

As we have stated in several places above, our work has caveats. A major one is that all calculations are carried out with the minimum model of neutron stars consisting of nucleons, electrons and muons only without treating the physics of new phases and new degrees of freedom that may appear. In particular, it is known that the appearance of Delta resonances and some hyperons is strongly correlated with the density dependence of nuclear symmetry energy, albeit depending also sensitively on their poorly known couplings/interactions with other particles  \cite{Cai,Drago}. Some of these particles may start appearing at densities around $(2-3)\rho_0$. It thus remains an interesting question how the appearance of these particles may affect the density dependence of the speed of sound as well as the onset densities or possible removals of the instabilities we studied in this work. \\

\section*{Acknowledgement}
This work is supported in part by the U.S. Department of Energy, Office of Science, under Award Number DE-SC0013702, the CUSTIPEN (China-U.S. Theory Institute for Physics with Exotic Nuclei) under the US Department of Energy Grant No. DE-SC0009971, the National Natural Science Foundation of China under Grant No. 12005118, and the Shandong Provincial Natural Science Foundation under Grants No. ZR2020QA085.
%
%

\end{document}